\documentclass[conference]{IEEEtran}
\IEEEoverridecommandlockouts
\usepackage{cite}
\usepackage{amsmath,amssymb,amsfonts,amsthm,csquotes}

\usepackage{algorithmic}
\usepackage{graphicx}
\usepackage{textcomp}
\usepackage{xcolor}
\usepackage{mathtools}
\usepackage{dsfont}

\theoremstyle{plain}
\newtheorem{theorem}{Theorem}

\newtheorem{proposition}{Proposition}
\newtheorem{corollary}{Corollary}

\theoremstyle{remark}

\theoremstyle{definition}

\def\BibTeX{{\rm B\kern-.05em{\sc i\kern-.025em b}\kern-.08em
    T\kern-.1667em\lower.7ex\hbox{E}\kern-.125emX}}
  
\begin{document}

\title{Random Access with Delayed Feedback: Stability, Delay, and Age of Information\\

\thanks{This work was supported in part by Turk Telekom (with grant no 136179 and a scholarship within the 5G and Beyond Joint
Graduate Support Programme coordinated by Information and Communication
Technologies Authority), and by the European Union (ERC Advanced
Grant 101122990-GO SPACE-ERC-2023-A.) Views and opinions expressed
are those of the authors only and do not necessarily reflect
those of the funding agencies.}
}



\author{
			Yiğit İnce and
			Elif Uysal, \emph{Fellow, IEEE}\\
			\textit{Communication Networks Research Group (CNG), EE Dept, METU, Ankara, Turkiye} \\
			
			\textit{yigit.ince@metu.edu.tr, uelif@metu.edu.tr}\vspace{-1em}
}
\maketitle

\begin{abstract}
We study random access for channels with long propagation delay. In such channels, waiting for feedback requires slotted ALOHA to use large slots, leading to reduced throughput, increased delay and Age of Information (AoI). Various schemes have previously been proposed to handle the propagation delay more effectively, such as the Interleaved Channel Policy (ICP) and the Transmit and Wait Policy (TWP). Extending ideas from framed slotted ALOHA to these schemes, we devise two new schemes, called Framed Interleaved Channel Policy (FICP) and Framed Transmit and Wait Policy (FTWP), to handle propagation delay. We compare ICP, TWP, FICP, and FTWP in terms of stability region, delay, and Normalized Average AoI (NAAoI). We find that all perform similar in terms of NAAoI; TWP and FTWP perform superior to others with respect to stability region and delay, FTWP requiring significantly less frequent feedback than TWP.
\end{abstract}


\section{Introduction}

Random access has gained renewed attention with the rise of Nonterrestrial Networking  (NTN), in particular in 3GPP Release 17 \cite{r17} and onwards. An important difference of NTN with terrestrial radio access is the sheer distance between base stations and user equipment. Consequently, NTN channels are characterized by propagation delays significantly larger than those found in terrestrial wireless links. Meanwhile, especially for short, sporadic data transmission (e.g., IoT, the transmission of preambles for reservation), random access stands out as the most efficient radio access approach. In particular, traditional slotted ALOHA and its variants are enjoying renewed popularity due to their simplicity and light protocol overhead in the NTN-IoT environment\cite{aloha,slottedaloha}. On the other hand, the need for an acknowledgment to be received within the slot leads to variants of slotted ALOHA used in NTN to use large slots. In the case of mild to moderate traffic conditions, this is a factor that can drastically decrease throughput, causing delay, and aging of the data. 

Schemes to mitigate feedback delay have been proposed in the previous literature \cite{int,copley}. In \cite{int}, Interleaved Channel Policy (ICP) is introduced, in which the feedback delay is mitigated by dividing the channel into interleaved channels; whereas in \cite{copley}, a more homogeneous scheme called Transmit and Wait Policy (TWP) is introduced. The following are our contributions in this paper:
\begin{itemize}
  \item We introduced Framed Interleaved Channel Policy (FICP) and Framed Transmit and Wait Policy (FTWP), which are the framed counterparts of ICP and TWP. FICP and FTWP depends on less frequent feedback than ICP and TWP.
  \item We studied the throughput, approximate average packet delay, and Normalized Average Age of Information (NAAoI) of ICP, TWP, FICP, and FTWP. We found that all perform similar in terms of NAAoI, and TWP and FTWP perform better than others with respect to throughput and delay.
\end{itemize}


\section{System Model}
\label{sec:model}
We consider a collision channel with propagation delay. There is a central receiver and $M$ uncoordinated users where time is slotted and indexed by $k$, whereby each transmission begins and ends within one slot. New packets may arrive at the users at the beginning of each slot. We denote the total arrival rate by $\lambda$ (Arrivals/Slot) and its estimate by $\hat\lambda$. The arrivals for different users are independent Bernoulli processes with arrival rate $\theta_j$ for user $j$ such that $\sum_{j=1}^M\theta_j=\lambda$. When a user has a packet to send, we call it a backlogged user and denote the number of backlogged users at the beginning of the $k^{th}$ slot as $N_k$. Each user can store at most one packet, which means that when a new packet arrives to a backlogged user, it replaces the one in the buffer. The estimate of $N_k$ that each user calculates using the collision feedback is denoted as $\hat N_k$. 

\begin{figure}[t]
\centering
\includegraphics[width=\columnwidth, trim=0 22.3cm 0 1cm, clip]{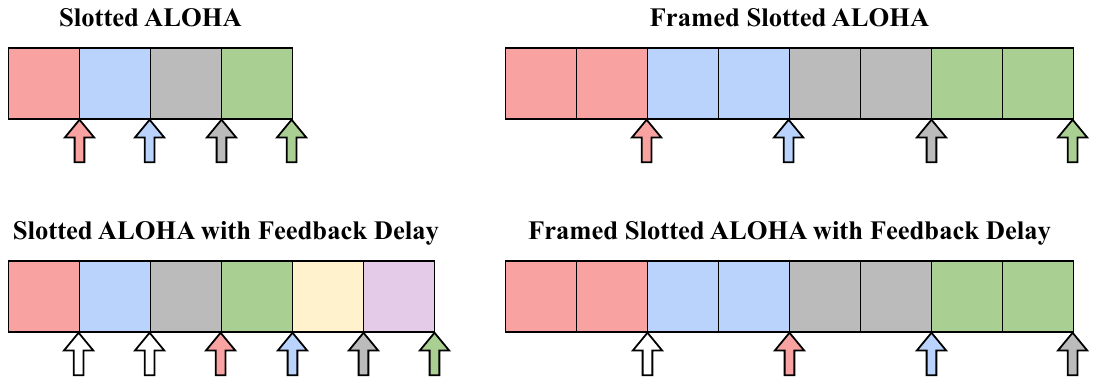}
\caption{Illustration of various random access schemes where arrows indicate the feedback for the slot or frame with the same color.}
\label{fig:delayedfsa}
\end{figure}


\section{Previous work on Handling Propagation Delay in Slotted ALOHA}
\label{sec:prev}

In slotted ALOHA, the feedback for the outcome of the slot is available at the end of each slot. In case the channel has a nonzero propagation delay, a possibility is to stay idle until the feedback is available to the users. This results in  slots that are larger than a transmission duration, reducing throughput. Alternatively, one can design a scheme that accommodates a feedback delay of a certain number of slots \cite{int,copley}. The schemes in \cite{int} and \cite{copley} will be the basis for our frame based analysis. In Fig. \ref{fig:delayedfsa}, feedback and frame structures are visualized.

\begin{figure}[t]
\centering
\includegraphics[width=\columnwidth]{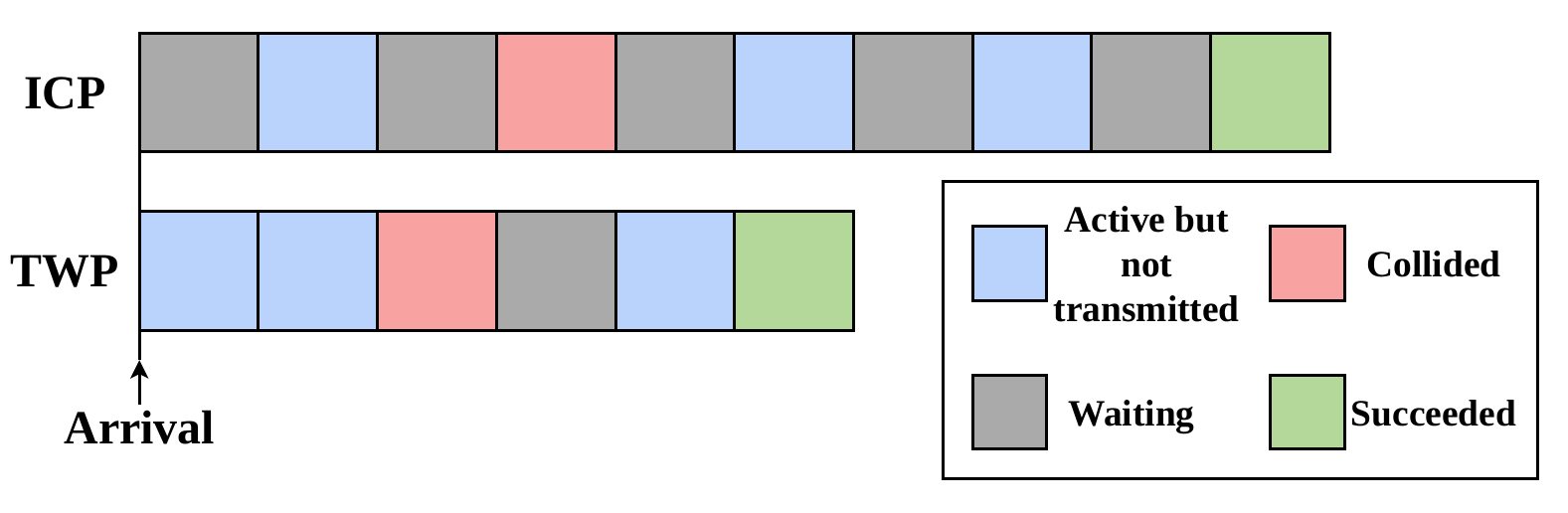}
\caption{Sample outcomes after an arrival until the first successful transmission for a user for ICP and TWP, $d_s=1$}
\label{fig:icp_twp}
\end{figure}

\subsection{The Interleaved Channel Policy (ICP)}

Given feedback delay of $d_s$ slots, the channel can be separated into $d_s+1$ interleaved channels where each interleaved channel behaves as an independent channel without feedback delay \cite{int}. An interleaved channel consists of slots with the same $\mathrm{mod}_{d_s+1}(k)$ value. Each user is assigned to one interleaved channel. This policy is a trade-off between Time Division Multiplexing (TDM) and random access. Groups of users are assigned to interleaved channels, but they still contend within the group. We denote the backlog in the $i^{th}$ interleaved channel at the beginning of the $k^{th}$ slot as $N^i_k$, and its estimate as $\hat N^i_k$. A backlogged user transmits with probability $1/\hat N^i_{k}$ in the $k^{th}$ slot if the $k^{th}$ slot is in the user’s interleave, and stays idle otherwise. The analysis of the ICP will require minimal effort since the interleaved channels behave as channels without feedback delay. Rivest's Pseudo Bayesian Estimator \cite{rivest} can directly be used here, for every interleaved channel $i$ there will be an estimate $\hat N^i_k$ of $N^i_k$
\begin{equation*}
\begin{aligned}
\hat N^i_{i+(k'+1)(d_s+1)} &= \max\{1, \hat N^i_{i+k'(d_s+1)} - (1-\mathds{1}_{\{col,\,k\}}) \\
&\quad + \tfrac{1}{e-2}\mathds{1}_{\{col,\,k\}} + \hat \lambda_i \},
\end{aligned}
\end{equation*}
where $\mathds{1}_{\{col,\,k\}}$ is the indicator function for the collision event in the $k=(i+k'(d_s+1))^{th}$ slot; and $\hat\lambda_i$ is the estimate of $\lambda_i$, the arrival rate for the $i^{th}$ interleaved channel.

\subsection{The Transmit and Wait Policy (TWP)}

In TWP, when a user transmits a packet, it waits for the feedback for $d_s$ slots \cite{copley}. If the transmission turns out to be a success, the user leaves the backlog. If the transmission turns out to result in a collision, the user is allowed to transmit again. The difference between TWP and ICP is as follows: Suppose there is a collision in the $0^{th}$ slot. Collided users wait for $d_s$ slots in both schemes and contend for the $(d_s+1)^{th}$ slot with the transmission probability determined by the backoff policy. The users collided in the $0^{th}$ slot but not transmitted in the $(d_s+1)^{th}$ slot can continue contending for the $(d_s+2)^{th}$ slot in TWP, whereas such users must wait for the $(2d_s+2)^{th}$ slot to continue contending in ICP. TWP does not have the interleaved channels or TDM structure that ICP has. The structure of TWP is more homogeneous than ICP. Sample outcomes after an arrival until the first successful transmission for a user for ICP and TWP can be seen in Fig. \ref{fig:icp_twp}. For TWP, since the feedback is available to the users after $d_s$ slots, the new estimate resulting from an outcome of a slot will be available to the users after $d_s$ time slots. Hence, for TWP, the estimate that is used to determine the transmission probabilities will lag by $d_s$ time slots behind the estimate that would be available in a channel without propagation delay. The estimator update equation is as follows \cite{rivest,copley}:
\begin{equation*}
\hat N_{k+1} =
\max\{1,\, \hat N_{k} - (1-\mathds{1}_{\{col,\,k\}})+ 
\frac{1}{e-2}\mathds{1}_{\{col,\,k\}}+ \hat \lambda \},  
\end{equation*}
where $\mathds{1}_{\{col,\,k\}}$ is the indicator function for the collision event in the $k^{th}$ slot.

Backlogged users transmit in slot $k$ with probability $1/\hat N_{k-d_s}$, as the most recent estimate available at the users at the beginning of the $k^{th}$ slot is $\hat N_{k-d_s}$.

\section{Framed Slotted ALOHA}
\label{sec:FSA}

Framed slotted ALOHA (FSA) refers to the schemes where slots are framed together such that the collective feedback for the frame is available at the end of the frame. A frame consists of $S$ slots. We call the users that are going to transmit in a frame attempting users. Attempting users choose a slot uniformly inside the frame. The first work where FSA is described is \cite{schoute}, where the main goal was to design an RFID scheme where arrivals are not stochastic, and
all backlogged users transmit in the frames with probability 1. 

In \cite{fpba}, the FSA scheme named Framed Pseudo-Bayesian ALOHA algorithm (FPBA), which we will utilize, is introduced. Frames are indexed by $t$, backlog at the beginning of the $t^{th}$ frame is denoted by $N_t$, and its estimate by $\hat N_t$. In a channel without propagation delay, backlogged users attempt to frame $t$ with probability $\min(S/\hat N_t,1)$.
$\hat N_t$ is updated at the end of the frame after the collision feedback is received by the users, according to a modified version of Rivest's Pseudo Bayesian Estimator \cite{rivest,fpba}:
\begin{equation}
\hat N_{t+1} =
\max\{1,\, \hat N_t - (S-s_{col,\,t})+ 
\frac{1}{e-2}s_{col,\,t}+ \hat \lambda S \},    
\end{equation}
where $s_{col,\,t}$ is the number of collisions in the $t^{th}$ frame.
The users that became backlogged during the frame can only attempt in the next frame.

We can calculate the throughput of FPBA, which is equal to the expected successful transmissions per slot. Since all slots are identical (not independent), we can find the expected successful transmissions for an arbitrary slot. The probability that an arbitrary backlogged user attempting an arbitrary slot in the $t^{th}$ frame is equal to $1/\max(\hat N_t,\,S)$. We define $N_t'$ as $\max(\hat N_t,\,S)$ for convenience. Thus, the successful transmission probability of an arbitrary user $j$ in an arbitrary slot $k$ in frame $t$ is
\begin{equation}
\mathbb{P}(\{\mathrm{Success}_j(k)\})=   \frac{1}{N_t'}\Big(\frac{N_t'-1}{N_t'}\Big)^{N_t-1}.
\end{equation}
Since the events $\{\mathrm{Success}_{j_1}(k)\}$ and $\{\mathrm{Success}_{j_2}(k)\}$ are disjoint for $j_1 \ne j_2$, we can write the total probability of a successful transmission in an arbitrary slot $k$ as follows:
\begin{equation}
\mathbb{P}(\{\mathrm{Success}(k)\})=   \frac{N_t}{N_t'}\Big(\frac{N_t'-1}{N_t'}\Big)^{N_t-1}.
\end{equation}
This expression is the same successful transmission probability expression with slotted ALOHA, except instead of $\hat N_t$, we have $N'_t=\max(\hat N_t,\,S)$. This means that FPBA has the same throughput expression as slotted ALOHA when $\hat N_t \ge S$.

\section{FSA for Channels with Propagation Delay}
\label{sec:FSAPD}
 Consider a channel with a feedback delay of $d_s$ slots. Define $d_f$ as the feedback delay in terms of frames. When the frame size is equal to $S$, $d_f$ is equal to $\lceil d_s/S \rceil$. For example, if the frame size $S$ is selected as $d_s$, $d_f$ becomes 1. We will propose two schemes, which we call Framed Interleaved Channel Policy (FICP) and Framed Transmit and Wait Policy (FTWP).
\subsection{Framed Interleaved Channel Policy (FICP)}

In ICP, an interleaved channel consists of slots with the same $\mathrm{mod}_{d_s+1}(k)$ value. We propose FICP, where frames are grouped together such that they form interleaved channels, an interleaved channel consisting of frames with the same $\mathrm{mod}_{d_f+1}(t)$ value.

In this policy, every interleaved channel behaves as a framed slotted ALOHA scheme without feedback delay. Users in different interleaved channels have different estimates of the backlog, as in ICP. We denote the backlog in the $i^{th}$ interleaved channel at the beginning of the $t^{th}$ frame as $N^i_t$, and its estimate as $\hat N^i_t$. The same backlog estimator as the one used for FPBA \cite{rivest,fpba} is employed:
\begin{equation}
\begin{aligned}
\hat N^i_{i+(t'+1)(d_f+1)} &= \max\{1, \hat N^i_{i+t'(d_f+1)} - (S-s^i_{col,\,t}) \\
&\quad + \tfrac{1}{e-2}s^i_{col,\,t} + \hat \lambda_i S \},
\end{aligned}
\end{equation}
where $s^i_{col,\,t}$ is the number of collisions in the $t=(i~+~t'~(d_f~+~1))^{th}$ frame.
A backlogged user transmits with probability $\mathrm{min}(1,S/\hat N^i_t)$ in the $t^{th}$ frame if this frame belongs to the user's interleaved channel, and stays idle otherwise.
\subsection{Framed Transmit and Wait Policy (FTWP)}

In TWP, users wait for the feedback for $d_s$ slots after transmission. If the transmission turns out successful, the user leaves the backlog; if the transmission turns out to result in a collision, the user is allowed to transmit again. We propose FTWP, where users wait for the feedback for $d_f$ frames after transmission. If the transmission turns out successful, the user leaves the backlog; if the transmission turns out to result in a collision, the user is allowed to transmit again as in TWP.

Since feedback is available at the users after $d_f$ frames, the new estimate resulting from the outcome of a frame will be available to the users after $d_f$ time frames. The estimate used to determine the transmission probabilities will lag by $d_f$ time frames behind the estimate that would be available in a channel without propagation delay. The estimator update equation is as follows \cite{rivest}:
\begin{equation}
\hat N_{t+1} =
\max\{1,\, \hat N_t - (S-s_{col,\,t})+ 
\frac{1}{e-2}s_{col,\,t}+ \hat \lambda S \}.    
\end{equation}

Backlogged users transmit in frame $t$ with probability $\mathrm{min}(1,S/\hat N_{t-d_f})$ using the most recent estimate available to them at the beginning of the $t^{th}$ frame, $\hat N_{t-d_f}$.
\section{Performance Comparisons}
\label{sec:comp}

We will compare ICP, TWP, FICP, and FTWP in terms of stability region, delay statistics, and Age of Information (AoI) for a channel with propagation delay.

\subsection{Stability Region Comparison}

In order to compare the stability regions, we take $M$ to infinity. As $M$ approaches infinity, the total arrivals in each slot become Poisson with mean $\lambda$. 
For TWP and FTWP, we consider a set of infinite users with buffer space for only one packet; and for ICP and FICP, we consider $d_s+1$ and $d_f+1$ sets of infinite users with one buffer, respectively.

\begin{theorem}
    \label{theorem:stability}
    TWP and FTWP are stable for $\lambda<e^{-1}$, $0<\hat\lambda\le e^{-1}$, and $\lambda \le \hat \lambda$.
\end{theorem}
\begin{proof}[Proof Sketch]
The stability can be proven by extending the Lyapunov function constructed by Tsitsiklis \cite{tsitsiklis} to a state space with $2(d_s+1)$ dimensions for TWP. The state vector $X_k$ consists of one variable for backlogged users allowed to transmit, $d_s$ variables for backlogged users waiting for feedback, and $d_s+1$ variables for the estimator. Similarly, for FTWP, a $2(d_f+1)$ dimensional state space is needed. The proof proceeds by showing that for sufficiently large backlog or estimation error, the drift of the Lyapunov function is negative. The key technical step lies in showing that the \enquote{restoring force} of the update algorithm (which scales with the estimation error) dominates the \enquote{noise} introduced by the delay, provided the delay $d_s$ is finite. This suggests that the Markov chain $\{X_k\}$ is geometrically ergodic. The detailed proof of stability will appear in the full version of this paper.  
\end{proof}

For ICP, there are $d_s+1$ interleaved channels where each interleaved channel has its own set of users. For ICP to be stable, all interleaved channels must be stable. An interleaved channel behaves as a slotted ALOHA channel without feedback delay, stable for arrival rates up to $\frac{e^{-1}}{d_s+1}$ \cite{tsitsiklis}, hence the arrival rate for each interleaved channel must satisfy
\begin{equation}
\label{eq:icpthroughput}
    \lambda_i<\frac{e^{-1}}{d_s+1}, \, i= 1,...,d_s+1.
\end{equation}

For FICP, there are $d_f+1$ interleaved channels where each behaves as an FSA scheme without feedback delay. Similar to ICP, for FICP to be stable, all interleaved channels must be stable.
An interleaved channel for FICP is stable if $\lambda_i<\frac{e^{-1}}{d_f+1}$ \cite{stabiFSA}, hence the arrival rate for each interleaved channel must satisfy
\begin{equation}
\label{eq:ficpthroughput}
\lambda_i<\frac{e^{-1}}{d_f+1}, \, i \in 1, ..., d_f+1.
\end{equation}
\begin{corollary}
\label{corollary:subset}
    The stability regions of ICP and FICP are subsets of the stability regions of TWP and FTWP.
\end{corollary}

\begin{proof}
For ICP, when the arrival rate for interleaved channel $i$ is $\lambda_i$, the total arrival rate $\lambda$ is equal to $\sum_{i=1}^{d_s+1} \lambda_i$. If the ICP scheme is stable, then $\lambda_i<\frac{e^{-1}}{d_s+1}, \, i \in 1,2,...,d_s+1$. Hence, $\sum_{i=1}^{d_s+1} \lambda_i=\lambda<e^{-1}$. Thus, the stability of ICP implies the stability of TWP and FTWP using Theorem \ref{theorem:stability}, but the converse is not true in general. The same argument can be made for FICP by substituting $d_s$ by $d_f$. 
\end{proof}

A $\gamma$-to-one mapping from the interleaved channels of ICP to the interleaved channels of FICP where $\gamma$ is a fixed number does not exist in general, hence we cannot claim that the stability region of ICP is a subset of the stability region of FICP and vice versa. Since $d_f$ is smaller than $d_s$, FICP has fewer interleaved channels, but more users and higher capacity for each interleaved channel. This suggests that FICP is more resilient to arrival bursts than ICP. A small burst may result in the arrival rate being several times the capacity of an interleaved channel when the interleaved channel has a small capacity as in ICP.

In terms of the stability region, TWP and FTWP are superior to ICP and FICP because of the unified channel structure. Now we will analyze the delay performances of these schemes.

\subsection{Approximate Delay Comparison}

In this section, we will analyze the expected delay in terms of slots for approximate models of TWP, FTWP, and ICP. FICP is skipped, as it is rather complex to analyze and may not give us any insight. We define the expected delay $\overline{D}_\pi$ as the expected time from the arrival of a packet until its delivery to the receiver for policy $\pi$. We will use the approximate delay analysis conducted by Bertsekas and Gallager \cite{DataNetworks}. In \cite{DataNetworks}, the stabilized slotted ALOHA scheme is modeled by an M/G/1 queue, with Poisson arrivals with rate $\lambda$, and departures happening at discrete time instants. The probability of successful transmission $P_{succ}$ is set as $e^{-1}$ for $N \ge 2$, and $P_{succ}=1$ for $N=1$, where $N$ is the backlog. 

\begin{proposition}\cite{DataNetworks}
\label{proposition:datanetworks}
    Let $W_b$ be the delay from the arrival of the $b^{th}$ packet until the beginning of the $b^{th}$ successful transmission. Note that if the system were first-come first-serve, $W_b$ would be the queuing time of the $b^{th}$ arrival. The average of $W_b$ over all $b$ is equal to the expected queuing delay $\overline W$. $W_b$ can be expressed as
    \begin{equation}
    \label{eq:datanetworks}
W_b=R_b+\sum_{c=1}^{n_b}\tau_c+y_b,
    \end{equation}
    where $R_b$ is the residual time to the beginning of the next slot; $\tau_1$ is the subsequent interval until the next successful transmission is completed; $\tau_c$, $1<c\le n_b$, is the interval from the end of the $(c-1)^{th}$ subsequent success to the end of the $c^{th}$ subsequent success; $n_b$ is the number of backlogged packets at the instant before $b's$ arrival, excluding current successful transmissions and including current unsuccessful transmissions. After those $n_b$ successes, $y_b$ is the remaining interval until the beginning of the $b^{th}$ successful transmission. 
\end{proposition}

Throughout this section, we will also assume that arrivals are a Poisson process as in \cite{DataNetworks} to make the analysis simpler. For ICP, we define a cycle which begins and ends at the beginning of every slot corresponding to the $i^{th}$ interleaved channel. A cycle consists of $d_s+1$ slots. There can be a departure only in the first slot of a cycle for the $i^{th}$ interleaved channel. Every interleaved channel of ICP behaves as a slotted ALOHA scheme without feedback delay, therefore we set $P_{succ}=e^{-1}$ for $N \ge 2$ and $P_{succ}=1$ for $N = 1$ as in \cite{DataNetworks}, where $N$ is the backlog at the beginning of every cycle.

\begin{corollary}
For the approximate model of ICP, the expected delay $\overline D_{ICP,i}$, from the arrival of a packet until its delivery to the receiver is
\begin{equation}
\label{eq:icp_delay}
\begin{aligned}
\overline D_{ICP,i} &= \frac{(e-\frac{1}{2})(d_s+1)-\frac{(e\lambda_i(d_s+1)-1)(e^{\lambda_i(d_s+1)}-1)}{(e^{\lambda_i(d_s+1)}-1-\frac{1}{e-1})\lambda_i}}{1-e\lambda_i(d_s+1)} \\
&\quad + 1 + \frac{d_s}{2}.
\end{aligned}
\end{equation}
\end{corollary}
\begin{proof}
  We can write $W_b$ as in (\ref{eq:datanetworks}),
 where $R_b$ is the residual time in terms of slots to the beginning of the next cycle; $\tau_1$ consists of the subsequent cycles up to and including the cycle in which the next successful transmission is completed; $\tau_c$, $1<c\le n_b$, is the interval from the end of the $(c-1)^{th}$ subsequent success up to and including the cycle in which the $c^{th}$ subsequent successful transmission is completed. After those cycles containing $n_b$ successes, $y_b$ is the remaining interval until the beginning of the $b^{th}$ successful transmission.
 
The expected residual time to the beginning of the next slot of the $i^{th}$ interleaved channel is
 $\frac{d_s+1}{2}$.
 For each interval $\tau_c$, the backlog is at least 2. Therefore, the expectation of $\tau_c$ is
 $(d_s+1)e$, and the expectation of $n_b$ is $\lambda_i \overline W$ using Little's Theorem. We can write
 \begin{equation}
     \overline W=\frac{d_s+1}{2}+\lambda_i(d_s+1)\overline W e+\mathbb{E}\left[y_b\right].
 \end{equation}
 In order to find $\mathbb{E}\left[y_b\right]$, we do a similar analysis with \cite{DataNetworks}. After the cycle containing the $n_b^{th}$ success, if the backlog is $1$ at that point, $y_b=0$. If the backlog is greater than $1$, then the expectation of $y_b$ is $(e-1)(d_s+1)$. Let $p_N$ be the steady-state probability that the backlog is $N$ at the beginning of a cycle. Following the logic in \cite{DataNetworks}, $p_1$ is the fraction of slots in the $i^{th}$ interleaved channel in which $N=1$ and a packet is successfully transmitted, whereas $\lambda_i(d_s+1)$ is the fraction of slots in the $i^{th}$ interleaved channel in which a packet is successfully transmitted. Hence, $\frac{p_1}{\lambda_i(d_s+1)}$ is the fraction of packets transmitted from state $N=1$ and $1~-~\frac{p_1}{\lambda_i(d_s+1)}$ is the fraction transmitted from higher $N$. Therefore,
\begin{equation}
\begin{aligned}
    \mathbb{E}\left[y_b\right] &= \left(1-\frac{p_1}{\lambda_i(d_s+1)}\right)(e-1)(d_s+1) \\
    &= \frac{(\lambda_i(d_s+1)-p_1)(e-1)}{\lambda_i}.
\end{aligned}
\end{equation}
Using the arguments above, we can also write
\begin{equation}
    \lambda_i(d_s+1)=p_1+\frac{1-p_0-p_1}{e}.
\end{equation}
The state $N=0$ can be entered from the state $N=0$ and $N=1$ if no arrivals occur in the cycle between; hence, we can write
\begin{equation}
    p_0=(p_0+p_1)e^{-\lambda_i(d_s+1)}.
\end{equation}
Solving for $p_1$, we obtain
\begin{equation}
    p_1=\frac{\lambda_i(d_s+1)-e^{-1}}{1-e^{-1}-\frac{e^{-1}}{e^{\lambda_i(d_s+1)}-1}}.
\end{equation}
Solving for $\overline W$ and adding the deterministic transmission delay $1$ and the propagation delay $\frac{d_s}{2}$ yields the expression in (\ref{eq:icp_delay}).
\end{proof}
For the case where the arrival rates for different interleaved channels are equal, $\lambda_i=\frac{\lambda}{d_s+1}$. In such a case, the expected delay becomes
\begin{equation}
\label{eq:delayicp}
    \overline D_{ICP,i} =\frac{(e-\frac{1}{2}-\frac{(e\lambda-1)(e^{\lambda}-1)}{(e^{\lambda}-1-\frac{1}{e-1})\lambda})(d_s+1)}{1-e\lambda}+1+\frac{d_s}{2}.
\end{equation}

For TWP, we can also use the model in \cite{DataNetworks}. We conservatively assume $P_{succ}=e^{-1}$ for all $N \ge 1$ to approximate the impact of stale backlog estimates. Note that these models are just approximate models as in \cite{DataNetworks}.

\begin{figure}[t]
\centering
\includegraphics[width=\columnwidth]{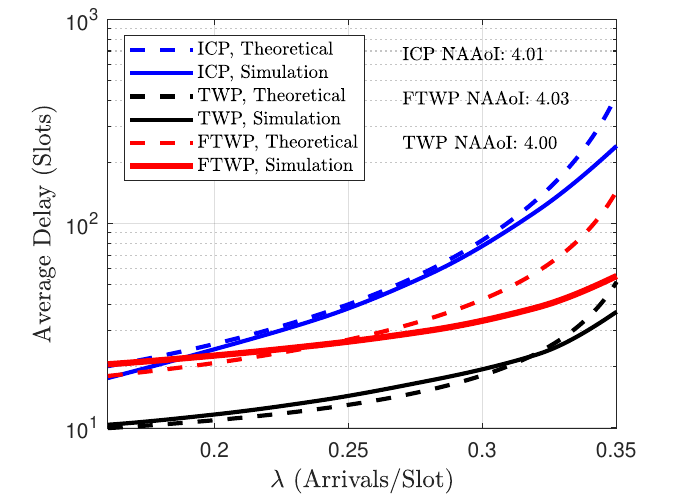}
\caption{Average delay expressions (\ref{eq:delayicp}), (\ref{eq:delaytwp}), and (\ref{eq:delayftwp}) compared with simulation results over a range of $\lambda$, together with simulation results for NAAoI for $\lambda=0.25$.
$d_s=10$ for all schemes and $S=10$ for FTWP. Simulated number of users $M$ is 1000 and the simulation duration is 300000 frames for all schemes.}
\label{fig:delay}
\end{figure}

\begin{corollary}
For the approximate model of TWP, the expected delay $\overline D_{TWP}$, from the arrival of a packet until its delivery to the receiver is
\begin{equation}
\label{eq:delaytwp}
    \overline D_{TWP}=\frac{2-2\lambda+e^{-1}}{2(e^{-1}-\lambda)}+\frac{d_s}{2}.
\end{equation}
\end{corollary}
\begin{proof}
We can write $W_b$ as a sum as in \cite{DataNetworks}
 \begin{equation}
        W_b=R_b+\sum_{c=1}^{n_b}\tau_c+y_b,
 \end{equation}
where $R_b$, $\tau_c$, $n_b$, and $y_b$ are defined the same as in Proposition \ref{proposition:datanetworks}. The expected residual time to the beginning of the next slot is $1/2$.
 The expectation of $\tau_c$ is
 $e$ and the expectation of $y_b$ is 
$e-1$. Using Little's Theorem, $\mathbb{E}\left[n_b\right]=\lambda \overline{W}$. Therefore, we can write
\begin{equation}
   \overline{W}=\frac{1}{2}+\lambda e \overline{W} +e-1.
\end{equation}
Solving $\overline{W}$ and adding the deterministic transmission delay $1$ and propagation delay $\frac{d_s}{2}$ yields the expression in (\ref{eq:delaytwp}).
\end{proof}
For FTWP, we will use a similar model to TWP. When a packet arrives, it must wait for the next frame instead of waiting only for the next slot. Instead of doing a frame-based analysis, we adopt a slot based analysis for convenience. This ignores frame-level batch service effects. We set $P_{succ}=e^{-1}$ for all $N \ge 1$ for every slot as in TWP.   
\begin{corollary}
For the approximate model of FTWP, the expected delay $\overline D_{FTWP}$, from the arrival of a packet until its delivery to the receiver is
\begin{equation}
\label{eq:delayftwp}
    \overline D_{FTWP}=\frac{Se^{-1}+2-2\lambda}{2(e^{-1}-\lambda)}+\frac{d_s}{2}.
\end{equation}
\end{corollary}
\begin{proof}
We can write $W_b$ as a sum as in \cite{DataNetworks}
 \begin{equation}
        W_b=R_b+\sum_{c=1}^{n_b}\tau_c+y_b.
 \end{equation}
The expected residual time to the beginning of the next frame is $S/2$.
 The expectation of $\tau_c$ is
 $e$ and the expectation of $y_b$ is 
$e-1$. Using Little's Theorem, $\mathbb{E}\left[n_b\right]=\lambda \overline{W}$. Therefore, we can write
\begin{equation}
   \overline{W}=\frac{S}{2}+\lambda e \overline{W} +e-1.
\end{equation}
Solving $\overline{W}$ and adding the deterministic transmission delay 1 and propagation delay $\frac{d_s}{2}$ yields the expression in (\ref{eq:delayftwp}).
\end{proof}
In Fig. \ref{fig:delay}, we compare the delay expressions (\ref{eq:delayicp}), (\ref{eq:delaytwp}), (\ref{eq:delayftwp}), and the simulation results. We observe that TWP has the best delay performance, and FTWP performs close to TWP for large arrival rates.

\subsection{Age of Information Comparison}

In this section, we will do a very basic Normalized Average Age of Information (NAAoI) analysis. We denote the total number of users by $M$. The arrivals are Bernoulli processes with parameter $\theta$ for any user. Hence, the total arrival rate $\lambda$ is equal to $M\theta$.

The NAAoI for policy $\pi$ is denoted by $J^{\pi}(M)$ and defined as follows in \cite{sat}:
\begin{equation}
\label{eq: NAAoI}
J^{\pi}(M) = \lim_{K \to \infty} \mathbb{E}\!\left[\frac{1}{MK} \sum_{j=1}^{M} \sum_{k=1}^{K} \frac{h^{\pi}_j(k)}{M}\right],
\end{equation}
where $K$ is the time horizon and $h^\pi_j(k)$ is the AoI of the $j^{th}$ user at the end of the $k^{th}$ slot. In \cite{sat}, Proposition \ref{proposition:sat} is proved.

\begin{proposition}\cite{sat}
\label{proposition:sat}
For any stabilized slotted ALOHA scheme $\pi$, if $\lambda=M\theta<\lim_{M \to \infty}C_{\pi}(M)$, then
    $\lim_{M \to \infty}J^{\pi}(M)~=~\lambda^{-1}$,
where $C_{\pi}(M)$ is the maximum throughput of policy $\pi$ with $M$ users.
\end{proposition}

For TWP and FTWP, we define the NAAoI as in (\ref{eq: NAAoI}) and denote them by $J^{TWP}(M)$ and $J^{FTWP}(M)$ respectively.

\begin{corollary}
\label{corollary:AgeTWPFTWP}
For TWP and FTWP, if $\lambda=M\theta<e^{-1}$, then
   \begin{equation}
   \lim_{M \to \infty}J^{TWP}(M)=\lim_{M \to \infty}J^{FTWP}(M)=\frac{1}{\lambda}.
   \end{equation}
\end{corollary}

\begin{proof}
 TWP and FTWP are stabilized slotted ALOHA schemes such that 
$\lim_{M \to \infty} C_{TWP}(M)=\lim_{M \to \infty} C_{FTWP}(M)=e^{-1}$, from Theorem \ref{theorem:stability}. Hence, using Proposition \ref{proposition:sat}, we can state Corollary \ref{corollary:AgeTWPFTWP}.
\end{proof}
For ICP and FICP, there are $d_s+1$ and $d_f+1$ independent channels with possibly different arrival rates $\lambda_i$ respectively. In the definition of NAAoI in (\ref{eq: NAAoI}), the average is taken among all users. In ICP and FICP, taking the average among users in the same interleaved channel and finding the NAAoI as a function of the interleaved channel $i$ can provide more insight. We define $\alpha_i$ as the ratio of the number of users in the $i^{th}$ interleaved channel to $M$, and we define $\theta_i$ as the arrival rate for any user in the $i^{th}$ interleaved channel. We define the NAAoI for policy $\pi \in \{\mathrm{ICP}, \mathrm{FICP}\}$ as follows:
\begin{equation}
J_i^{\pi}(M)
= \lim_{K \to \infty} \mathbb{E}\!\left[
\frac{1}{MK}
\sum_{j=1}^{\alpha_i M}
\sum_{k=1}^{K}
\frac{h^{\pi}_j(k)}{\alpha_i M}
\right].
\end{equation}
Note that for ICP and FICP, we calculate the average among users in the $i^{th}$ interleaved channel, but normalize it by the total number of users, $M$, to ensure a fair comparison between policies based on TWP and ICP.

\begin{corollary}
\label{corollary:AgeICPFICP}
For ICP, if $\lambda_i=\alpha_iM\theta_i<\frac{e^{-1}}{d_s+1}$, then
   \begin{equation}
   \lim_{M \to \infty}J_i^{ICP}(M)=\frac{\alpha_i}{\lambda_i}.
   \end{equation}
Similarly for FICP, if $\lambda_i=\alpha_iM\theta_i<\frac{e^{-1}}{d_f+1}$, then
   \begin{equation}
   \lim_{M \to \infty}J_i^{FICP}(M)=\frac{\alpha_i}{\lambda_i}.
   \end{equation}
\end{corollary}

\begin{proof}
Every interleaved channel of ICP is a stabilized slotted ALOHA scheme such that $\lim_{M \to \infty} C_{i}(\alpha_iM)=\frac{e^{-1}}{d_s+1}$ from (\ref{eq:icpthroughput}). Using Proposition \ref{proposition:sat} and given the arrival rate condition for ICP in Corollary \ref{corollary:AgeICPFICP}, we can write
\begin{equation*}
\begin{aligned}
&\lim_{M \to \infty}J_i^{ICP}(M) = \lim_{M \to \infty}\lim_{K \to \infty} \mathbb{E}\!\left[ \frac{1}{MK} \sum_{j=1}^{\alpha_iM} \sum_{k=1}^{K} \frac{h^{\pi}_j(k)}{\alpha_iM}\right] \\
&= \alpha_i\lim_{M \to \infty}\lim_{K \to \infty} \mathbb{E}\!\left[ \frac{1}{\alpha_iMK} \sum_{j=1}^{\alpha_iM} \sum_{k=1}^{K} \frac{h^{\pi}_j(k)}{\alpha_iM}\right] \\
&=\alpha_i \lim_{M \to \infty}J^{\pi}(\alpha_iM)=\frac{\alpha_i}{\lambda_i}.
\end{aligned}
\end{equation*}
This completes the proof for ICP. The proof for FICP is the same except for the arrival rate condition.
\end{proof}
For the special case of $\alpha_i=\frac{1}{d_s+1}$ and $\lambda_i=\frac{\lambda}{d_s+1}$ for ICP, and $\alpha_i=\frac{1}{d_f+1}$ and $\lambda_i=\frac{\lambda}{d_f+1}$ for FICP, the NAAoI of TWP, FTWP, and any interleaved channel of ICP and FICP becomes the same. Simulation results for NAAoI for $\lambda=0.25$ are given in Fig. \ref{fig:delay}.
\section{Conclusion}
In this paper, we introduced FICP and FTWP, which are built upon ICP and TWP. We compared ICP, TWP, FICP, and FTWP in terms of stability region, average delay, and NAAoI. We showed that TWP and FTWP perform superior to the other two with respect to stability region and delay. We maintain that FTWP is a suitable choice as a random access scheme for satellite connections, as it achieves the best overall performance together while requiring much less frequent feedback than TWP.
\section{Acknowledgment}
The authors thank Aimin Li and Çağrı Arı for their valuable discussions and feedback on this paper.

\bibliographystyle{IEEEtran}
\bibliography{ref}

@article{tsitsiklis,
  author={Tsitsiklis, J.},
  journal={IEEE Transactions on Automatic Control}, 
  title={Analysis of a multiaccess control scheme}, 
  year={1987},
  volume={32},
  number={11},
  pages={1017-1020},
  doi={10.1109/TAC.1987.1104494}}

@article{rivest,
  author={Rivest, R.},
  journal={IEEE Transactions on Information Theory}, 
  title={Network control by Bayesian broadcast}, 
  year={1987},
  volume={33},
  number={3},
  pages={323-328},
  doi={10.1109/TIT.1987.1057315}}

@article{sat,
  author={Chen, Xingran and Gatsis, Konstantinos and Hassani, Hamed and Bidokhti, Shirin Saeedi},
  journal={IEEE Transactions on Information Theory}, 
  title={Age of Information in Random Access Channels}, 
  year={2022},
  volume={68},
  number={10},
  pages={6548-6568},
  doi={10.1109/TIT.2022.3180965}}

@article{stabiFSA,
  author={Yu, Jihong and Zhang, Pengfei and Chen, Lin and Liu, Jiangchuan and Zhang, Rongrong and Wang, Kehao and An, Jianping},
  journal={IEEE Journal on Selected Areas in Communications}, 
  title={Stabilizing Frame Slotted Aloha-Based IoT Systems: A Geometric Ergodicity Perspective}, 
  year={2021},
  volume={39},
  number={3},
  pages={714-725},
  doi={10.1109/JSAC.2020.3018795}}

@article{schoute,
  author={Schoute, F.},
  journal={IEEE Transactions on Communications}, 
  title={Dynamic Frame Length ALOHA}, 
  year={1983},
  volume={31},
  number={4},
  pages={565-568},
  doi={10.1109/TCOM.1983.1095854}}

@techreport{int,
  author      = {Georgiopoulos, M. and Merakos, L. and Papantoni-Kazakos, P.},
  title       = {An Efficient Random Access Algorithm for Packet Broadcast Channels with Long Propagation Delays},
  institution = {Dept of ECE, U. ofConnecticut},
  number      = {TR-85-9},
  address     = {Storrs, CT, USA},
  month       = nov,
  year        = {1985}}

@mastersthesis{copley,
  author       = {Copley, Jonathon H.},
  title        = {Random Access Satellite Communication in the Presence of Interference},
  school       = {Massachusetts Institute of Technology},
  department   = {Department of Electrical Engineering and Computer Science},
  type         = {Master's thesis},
  address      = {Cambridge, MA, USA},
  month        = may,
  year         = {2024}}

@article{fpba,
  author  = {Habachi, M. H. and Ali, B. M.},
  title   = {Adaptive Framed Pseudo-Bayesian Aloha Algorithm with Priorities},
  journal = {Electronics Letters},
  volume  = {36},
  number  = {13},
  pages   = {1145--1147},
  year    = {2000},
  month   = jun,
  doi     = {10.1049/el:20000845}
}

@techreport{r17,
  author      = {{3rd Generation Partnership Project (3GPP)}},
  title       = {{TR 21.917: Release 17 Description; Summary of Rel-17 Work Items}},
  institution = {3rd Generation Partnership Project (3GPP)},
  type        = {Technical Report},
  number      = {TR 21.917 V17.0.1},
  year        = {2022},
  note        = {Release 17}
}

@inproceedings{aloha,
  title={THE ALOHA SYSTEM: another alternative for computer communications},
  author={Abramson, Norman},
  booktitle={Proceedings of the November 17-19, 1970, fall joint computer conference},
  series={AFIPS '70 (Fall)},
  pages={281--285},
  year={1970},
  publisher={ACM},
  address={Houston, Texas},
  doi={10.1145/1478462.1478502}
}

@article{slottedaloha,
  title={ALOHA packet system with and without slots and capture},
  author={Roberts, Lawrence G},
  journal={Computer Communication Review},
  volume={5},
  number={2},
  pages={28--42},
  year={1975},
  publisher={ACM New York, NY, USA},
  doi={10.1145/1024916.1024920}
}

@book{DataNetworks,
  author    = {Bertsekas, Dimitri P. and Gallager, Robert G.},
  title     = {Data Networks},
  publisher = {Prentice Hall},
  address   = {Englewood Cliffs, NJ, USA},
  edition   = {2nd},
  year      = {1992}
}


\end{document}